# Submitted to Magnetic Resonance in Medicine

**High-Permittivity Pad Design Tool for 7T Neuroimaging and 3T Body Imaging**

**Author list:** Jeroen van Gemert[1], Wyger Brink[2], Andrew Webb[2], Rob Remis[1]

**Corresponding author:**

J.H.F. van Gemert

Circuits and Systems Group, Electrical Engineering, Mathematics and Computer Science Faculty

Delft University of Technology

Mekelweg 4, 2628 CD Delft, The Netherlands

Email: J.H.F.vanGemert-1@tudelft.nl

**Word count: 2596**

**Institution information:**

[1]Circuits and Systems Group, Electrical Engineering, Mathematics and Computer Science Faculty

Delft University of Technology

Mekelweg 4

2628 CD Delft

The Netherlands

[2]Radiology, C.J. Gorter Center for High-Field MRI,

Leiden University Medical Center

Albinusdreef 2

2333 ZA Leiden

The Netherlands

**Running title:** High-Permittivity Pad Design Tool

**Abstract**

**Purpose:** High-permittivity materials in the form of flexible "dielectric pads" have proved very useful for addressing RF inhomogeneities in high field MRI. Finding the optimal design of such pads is however a tedious task, reducing the impact of this technique. In this work we present an easy to use software tool which allows researchers and clinicians to design dielectric pads efficiently on standard computer systems, for 7T neuroimaging and 3T body imaging applications.

**Methods:** The tool incorporates advanced computational methods based on field decomposition and model order reduction as a framework to efficiently evaluate the $B_1^+$ fields resulting from dielectric pads. The tool further incorporates an optimization routine to perform either straightforward design for one or two pads based on a target field approach, or a trade-off analysis between homogeneity and efficiency of the $B_1^+$ field in a specific region of interest. The 3T version further allows for shifting of the imaging landmark to enable different imaging targets to be centered in the body coil.

**Results:** Example design results are shown for imaging the inner ear at 7T and for cardiac imaging at 3T. Computation times for all cases were under a couple of minutes. The change in homogeneity and efficiency with the pad's dimensions, location, and constitution are clearly listed.

**Conclusion:** The developed tool can be easily used to design pads for any 7T neuroimaging and 3T body imaging application within minutes. This bridges the gap between the advanced design methods and the practical application by the MR community.

**Introduction**

Obtaining MR images with spatially-invariant tissue contrast becomes more challenging at higher static magnetic field strengths. The fundamental reason for this is the increase in Larmor frequency, which leads to a shortened wavelength of the RF field. For static fields strengths of 3T and higher, this wavelength becomes comparable to the dimensions of the body, or shorter. As a consequence, wave-interference effects become apparent that reduce the homogeneity and strength of the transmit magnetic field, referred to as the $B_1^+$ field (1,2). The homogeneity of this field is of crucial importance in obtaining a uniform contrast in MRI.

Over the last decade, many RF shimming studies have been devoted to improving the $B_1^+$ field distribution and efficiency. Amongst popular methods are the active shimming techniques, where multiple separate transmit coils are used. The amplitudes and phases are configured for each element individually, such that the $B_1^+$ field is tailored in a certain region of interest (ROI) (3–6). Alternatively, dielectric materials can also be used to tailor the $B_1^+$ field, as a passive shimming approach. These materials typically have a high relative permittivity in the order of 80-300, and they induce a strong secondary magnetic field in their vicinity (7–15). The materials can be made fairly easy through aqueous suspensions of calcium titanate and/or barium titanate to obtain the appropriate permittivity (16–18). Subsequently, the mixture is sealed in a polypropylene bag with appropriate dimensions to form flexible pads. Typically, these dielectric pads are placed in close vicinity to the ROI tangent to the body.

Despite the ease of constructing such dielectric pads, their design is not trivial as it depends on many aspects; the optimal pad design varies with ROI, dimensions of the body, and MR configuration (e.g. static field strength and transmit antenna). Therefore, the pad's dimensions, location, and constitution need be optimized in an application-specific manner. One common approach is to perform a parametric optimization using general-purpose electromagnetic field solvers, based on a systematic trial-and-error approach and guided by user intuition, and then to choose the best pad-properties afterwards. As each of these simulations involve a large computational domain with an RF coil and heterogeneous body model, such procedures typically take multiple days for a single application (8,19–21). Some applications also benefit from having more than one dielectric pad, which further complicates the design procedure. This limits the exploitation of this practical shimming approach.

In previous work (22), we have developed an advanced reduced order modelling technique to accelerate pad evaluations by characterizing stationary components such as the RF coil and body model in an offline-stage, and compressing the resulting model. This yielded up to four orders of magnitude of acceleration when compared to using commercial software and enabled the automated design of a single dielectric pad in under a minute. Although these methods have been demonstrated, the offline procedures present an intractable task for any MR user planning to use dielectric pads due to either lack of software, resources, or expertise in this specific field. The approach also did not allow for designing two dielectrics at once, which can be beneficial in many applications.

In this work, we aim to bridge the gap between the advanced design methods and the practical application by the MR community, by integrating the automated design procedure in a stand-alone software tool. This tool can be run on a standard PC, is fast, and can be used to design multiple dielectric pads to optimize either the homogeneity or the efficiency of the $B_1^+$ field, or a combination of both, in any arbitrary ROI in the head at 7T or the body at 3T. Furthermore, for 3T, the imaging landmark of the transmit body coil can be shifted throughout the torso.

This note is structured as follows. The underlying methods of the tool are firstly set out in the methods section. Then the reduced order modeling and the optimization scheme are explained. Subsequently, the design tool is described which is then demonstrated cardiac imaging at 3T and for imaging the inner ear at 7T.

## Methods

### Configuration

The neuroimaging configuration was simulated using a shielded and tuned high-pass birdcage head coil with a radius of 15 cm operating at 298 MHz (7T). The body model "Duke" from the Virtual Family dataset was used (23), and the computational domain was uniformly discretized on a uniform 5 mm3 grid. The pad-design domain was taken as a 1 cm thick layer around the head model, which is constrained in practice by the tight-fitting receive arrays employed in MR.

The body imaging configuration was simulated using a wide-bore high-pass birdcage body coil with a radius of 35 cm operating at 128 MHz (3T), where also the body model Duke is used. The computational domain was discretized on a uniform 7.5 mm3 grid and the pad-design domain was defined as a 1.5 cm thick layer around the torso extending from just below the top of the shoulders down to the hips. Whereas the position of the head with respect to the head coil is fixed in the 7T neuroimaging setting, this is not the case for 3T body imaging. Therefore, additional field simulations were performed for a 1.5 cm spaced range of imaging landmarks within the torso to enable shifting of the body coil for different body imaging applications.

All field quantities were normalized to 1 W input power.

*Modeling dielectrics*

The backbone of the design tool is a diakoptic modeling approach which stems from the work established in (24,25), where an efficient forward model was presented for evaluating the effect of a dielectric pad. The basic idea is to split the computational domain into two parts as illustrated in Figure 1a and 1b for 7T neuroimaging and 3T body-imaging, respectively. The first domain is stationary and consists of the heterogeneous body model and RF transmit coil. These components remain unaffected throughout the pad-simulations and can therefore be characterized in advance. The second domain is dynamic and confines the pad-design domain where any desired dielectric pad can be positioned during the design process, i.e. with arbitrary geometry, location, and material properties (possibly heterogeneous). This domain-splitting allows us to compute the pad-independent background fields and field response library in an offline-stage, such that only the pad-specific secondary field involving the pad-design domain needs to be computed in the online-stage. As this latter domain is much smaller than the original full computational domain, computations are accelerated without compromising accuracy. More details on this procedure can be found in (24,25).

The complexity of the calculations can be reduced further through the application of reduced order modeling techniques as has been shown in (22). In this procedure the practical degrees-of-freedom of the pad design problem (i.e. many fewer than allowed by the computational grid) are exploited to compress the model while preserving the essential properties of the model. To this end, the pad design is parametrized in terms of its width, height, location, and constitution, through the parameter vector $\mathbf{p} = [\varepsilon; z_T; z_B; \phi_L; \phi_R]$ as illustrated in Figure 1c. Subsequently, the forward model is compressed by projecting onto a reduced-order basis to further accelerate $B_1^+$ field computations for any arbitrary dielectric pad in under a second of computation time.

In the 3T configuration each landmark position of the body coil involves different background fields, due to the different positioning of the body model with respect to the birdcage. However, the field response library remains unchanged as this involves only the response of the pad design domain.

*Optimization*

The pad optimization problem is formulated using a target field approach, in which we aim to achieve a certain desired $B_1^+$ field magnitude in a defined ROI, which is achieved by minimizing a cost functional $C$ as a function of the pad-parameter vector $\mathbf{p}$. This functional is defined as

$$C(\mathbf{p}) = \frac{1}{2} \frac{\left\| B_1^+(\mathbf{p}) - B_1^{+;\text{desired}} \right\|_2^2}{\left\| B_1^{+;\text{desired}} \right\|_2^2}, \qquad (1)$$

where $B_1^{+;\text{desired}}$ is the desired $B_1^+$ magnitude in an ROI, $B_1^+(\mathbf{p})$ is the field due to a pad with model parameters $\mathbf{p}$. The cost function in Eq. (1) aims to minimize the discrepancy between the prescribed $B_1^+$ field and the $B_1^+$ field generated by the model, integrated over the ROI. To measure the quality of the resulting $B_1^+$ field, we evaluate its mean as a measure of transmit efficiency, and its coefficient of variation $C_v$ as measure of homogeneity. The latter is defined as the ratio of the standard deviation to the mean.

To minimize this nonlinear function, we use a gradient descent method combined with a line search to determine the step-size, as the gradient of the function can be computed analytically (24,26). Other methods can also be used, but we found that this is the most efficient and stable method in the context of this application. To increase stability and to limit undesired behavior, we adopted the following rules for the update steps:

1. For iteration $j$ compute gradient as $\mathbf{g}^j = \nabla_{\mathbf{p}} C$
2. Set update direction as

$$\mathbf{u}^j(1) = -\text{sign}\{\mathbf{g}^j(1)\} * 20 \qquad \text{\% } \varepsilon \text{ update}$$

$$\mathbf{u}^j(2) = -\text{sign}\{\mathbf{g}^j(2)\} * w_1 * 0.75 \text{ cm} \qquad \text{\% } z_T \text{ update}$$

$$\mathbf{u}^j(3) = -\text{sign}\{\mathbf{g}^j(3)\} * \frac{1}{w_1} * 0.75 \text{ cm} \qquad \text{\% } z_B \text{ update}$$

$$\mathbf{u}^j(4) = -\text{sign}\{\mathbf{g}^j(4)\} * w_2 * \frac{\pi}{8} \text{ cm} \qquad \text{\% } \phi_L \text{ update}$$

$$\mathbf{u}^j(5) = -\text{sign}\{\mathbf{g}^j(5)\} * \frac{1}{w_2} * \frac{\pi}{8} \text{cm} \qquad \text{\% } \phi_R \text{ update}$$

With the weights defined as

$$w_1 = \frac{|\mathbf{g}^j(2)|}{|\mathbf{g}^j(3)|} \text{ and } w_2 = \frac{|\mathbf{g}^j(4)|}{|\mathbf{g}^j(5)|}$$

and restricted to

$$\frac{1}{1.5} \leq w_1 \leq 1.5 \text{ and } \frac{1}{1.5} \leq w_2 \leq 1.5$$

3. Update pad parameters $\mathbf{p}$ as

$$\mathbf{p}^{j+1} = \mathbf{p}^j + \alpha\mathbf{u} \text{ with an optimum step-size } 0 \leq \alpha \leq 1 \text{ found by line search}$$

The weights $w_1$ and $w_2$ control the weight between related variables and serve to include gradient sensitive information in the update steps.

*Design tool*

The tool is implemented in MATLAB (R2015a, The MathWorks, Inc., Natick, Massachusetts, USA). It allows for computations on a GPU when available to speed up computations and requires approximately 3 GB of working memory for 7T neuroimaging and 7 GB for 3T body imaging. When the GPU is being used, it requires approximately 0.8 GB and 2.5 GB video memory for neuroimaging and body imaging, respectively. The tool is available for download as an executable file at https://paddesigntool.sourceforge.io.

The graphical user interface of the design tool is shown in Figure 2. The top row depicts the tissue map of the head for the transverse, coronal, and sagittal slice, as well as a 3D view of the slices. If desired, the $B_1^+$ fields without dielectrics can be shown here instead. The bottom row shows the $B_1^+$ field with pads, as well as a 3D view of the current pad design, which is updated throughout the optimization process. The contrast and brightness of all displays can be manually adjusted.

The imaging landmark of the 3T body coil can be chosen as is depicted in Figure 3. Afterwards, the electromagnetic fields are updated for the new imaging landmark.

The tool allows optimization of the $B_1^+$ field using either one or two dielectric pads. In view of its implementation we can limit the maximum allowed relative permittivity $\varepsilon_r$ and incorporate a realistic conductivity throughout the iterations. Furthermore, the desired $B_1^+$ field magnitude for the ROI can be set in $\mu T/\sqrt{W}$, or alternatively, a sweep can be executed over a discrete set of predefined target fields to enable a trade-off analysis between transmit efficiency and homogeneity.

The user can specify a 3D volume of interest, by drawing three ellipsoidal ROIs in the three isometric views. Alternatively, as an example, a predefined ROI can be selected from a list. Subsequently, the minimization can be carried out, during which the geometry of the dielectric pad is continuously updated and shown in the bottom-right corner.

After the minimization is complete, the results will be summarized in a separate window. The results display the dimensions (width, height, and thickness) and the dielectric properties of the optimized dielectric pad. Furthermore, the resulting transmit efficiency and coefficient of variation are listed for the scenarios with and without a dielectric pad. All results can be stored for later reference.

**Results**

For demonstration purposes, we show two examples in which the $B_1^+$ field is optimized in terms of homogeneity and magnitude: first in cardiac imaging at 3T using a single dielectric pad, and then in imaging the inner ear at 7T using two dielectric pads. All computations have been executed on an Intel Xeon CPU X5660 @ 2.80 GHz (dual core) equipped with a NVIDIA Tesla K40c GPU.

For the 3T application example the body coil is shifted to be centered at the heart, which is then assigned as the ROI. A sweep over a set of eight target fields is carried out using the sweep-option, which took less than 1.5 minutes to compute. The electrical conductivity of the pad was fixed at 0.2 S/m. The results for this sweep are shown in Figure 3a. From the $B_1^+$-$C_v$ curvature we choose iteration number 3 to be the optimal pad design, a dielectric with dimensions 20 x 28 x 1.5 cm$^3$ and a relative permittivity of 213. This design corresponds to a target field equal to 0.51 μT/√W, as it improved the transmit efficiency by 23% and reduced the $C_v$ from 13.3% to 6.4%. The optimization results are illustrated in Figure 3b and the result corresponds with findings from (19). Higher efficiencies can be obtained as well, e.g. iteration number 5, but as the dimensions of the dielectric pad are 30 x 73 x 1.5 cm$^3$ it becomes less practical.

For the 7T inner ear application example an ROI is drawn that covers both inner ears, and the $B_1^+$ field is optimized with two dielectrics pads. The sweep-option identified the optimal pad design in under 2 minutes, which increased the transmit efficiency by 46% from 0.31 μT/√W to 0.45 μT/√W and decreased the $C_v$ from 38% to 13%. The corresponding $B_1^+$ field and design summary are shown in Figure 4a and 4b, which suggested two pads with dimensions 11 x 16 x 1 cm$^3$ and a relative permittivity of 269, and the second one with dimensions 22 x 12 x 1 cm$^3$ and a relative permittivity of 300, which agrees with previous findings (21).

**Discussion and Conclusions**

In this work we have presented a software tool that allows for designing dielectric pads for an arbitrary ROI in 7T neuroimaging and 3T body imaging applications. Computations are fast due to the underlying reduced order model, which enables MR operators to identify the optimal pad geometry and constitution in a matter of minutes. Aided by an optimization scheme, the optimal geometry, location, and material properties of the pad can be found which improve the $B_1^+$ field magnitude or homogeneity within the ROI, or a combination of both.

The optimization method used in the tool has been chosen in view of its stability. Other methods may be considered as well, such as Gauss-Newton methods which incorporate an approximant of the Hessian in the gradient direction. In our case, the Hessian is rank deficient and hence would need to be regularized in an application-specific manner, so this approach was not pursued here. In addition, we truncated the update steps to be limited to a certain range. For example, relative permittivity updates were forced to be less than 20, and updates of the top-end of the dielectric were forced to be less than 0.75 cm. The exact update is determined by the step size that is determined by the line search. Without these limitations, the algorithm had the tendency to converge to very large dielectrics with a low permittivity, which is not very practical to use. Using the truncated update steps, we avoided this undesired behavior.

The reduced order model that was used for 3T body imaging was created using a single birdcage landmark, centered at the liver. In the design tool the same model can be used to model pad-responses in other landmarks as well, without the need to compute this time-intensive part again. We found that the errors that are introduced by this approximation are minimal and do not affect the solution quality. This shows that a single library can be efficiently used for other applications as well, e.g. in case of using a local transmit coil or transmit array.

**Acknowledgments**

This project was funded by the Dutch Technology Foundation (STW) project 13375 and the European Research Council Advanced Grant 670629 NOMA MRI.

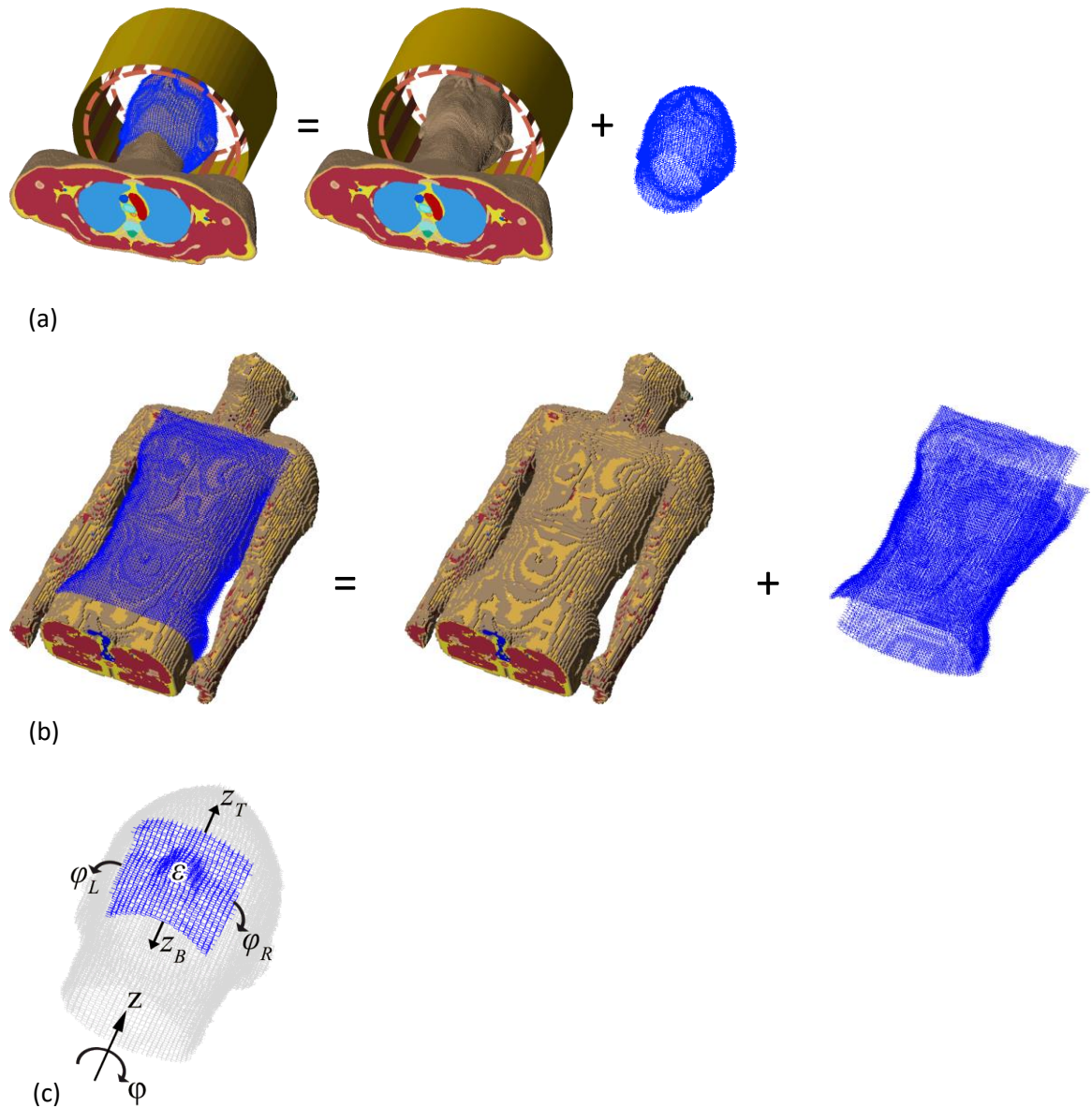

(a)

(b)

(c)

**Figure 1**. Splitting of computational domains and the parameterization of the dielectric pad. In **(a)** the 7T neuroimaging configuration is divided into a static part, consisting of a heterogeneous body model, RF coils, and an RF shield, and a dynamic part to which the dielectric is confined. In **(b)** the 3T body imaging configuration is shown, here the wide-bore birdcage is omitted for visualization purposes of the pad design domain. As only the pads that can be easily fabricated are of interest, the model is parameterized in the pad's characteristics as is shown in **(c)**.

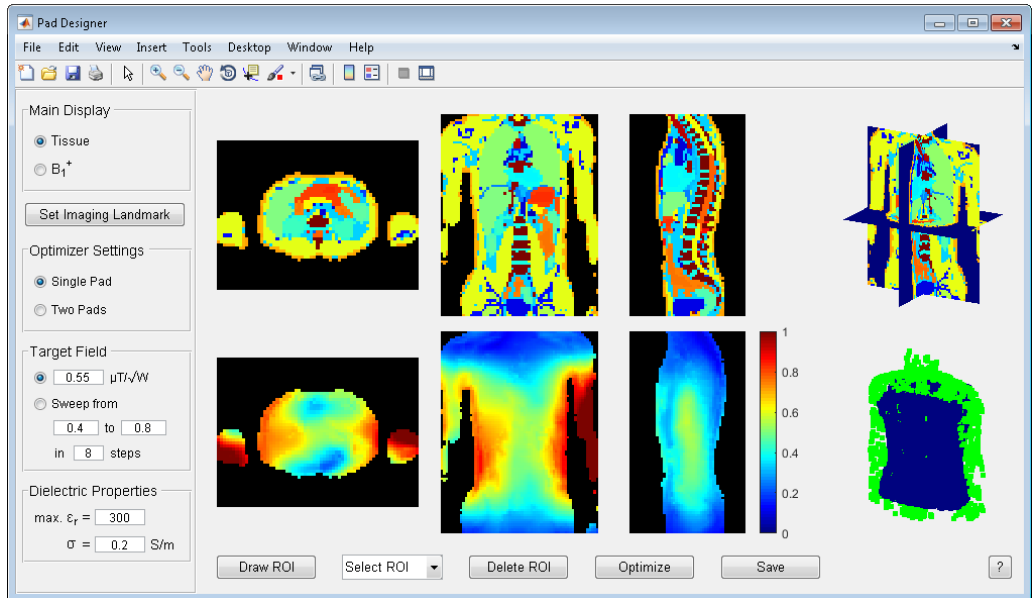

**Figure 2.** Graphical user interface of the pad design tool. Tissue profiles are shown in the top row, whereas $B_1^+$ fields are depicted in the bottom row. After a ROI is drawn, the user can start the optimization with the selected options.

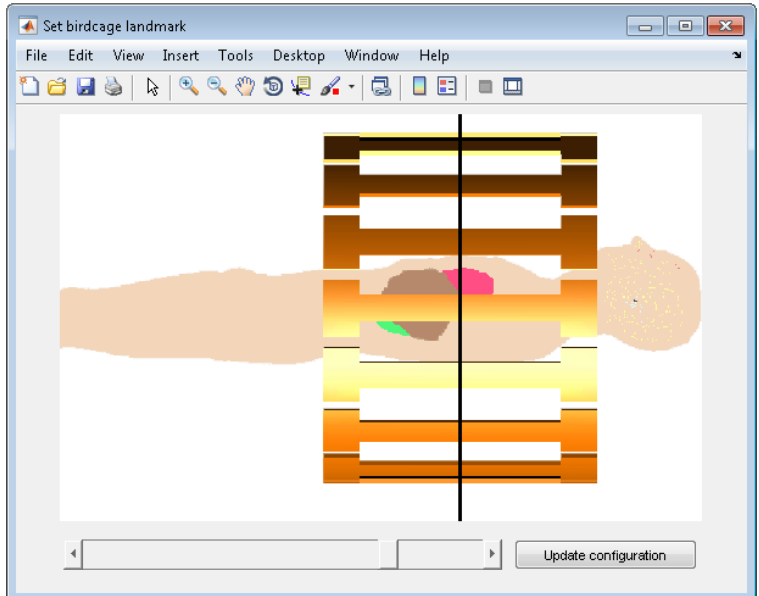

**Figure 3.** The center of the birdcage can be selected, after which the corresponding electromagnetic fields are updated.

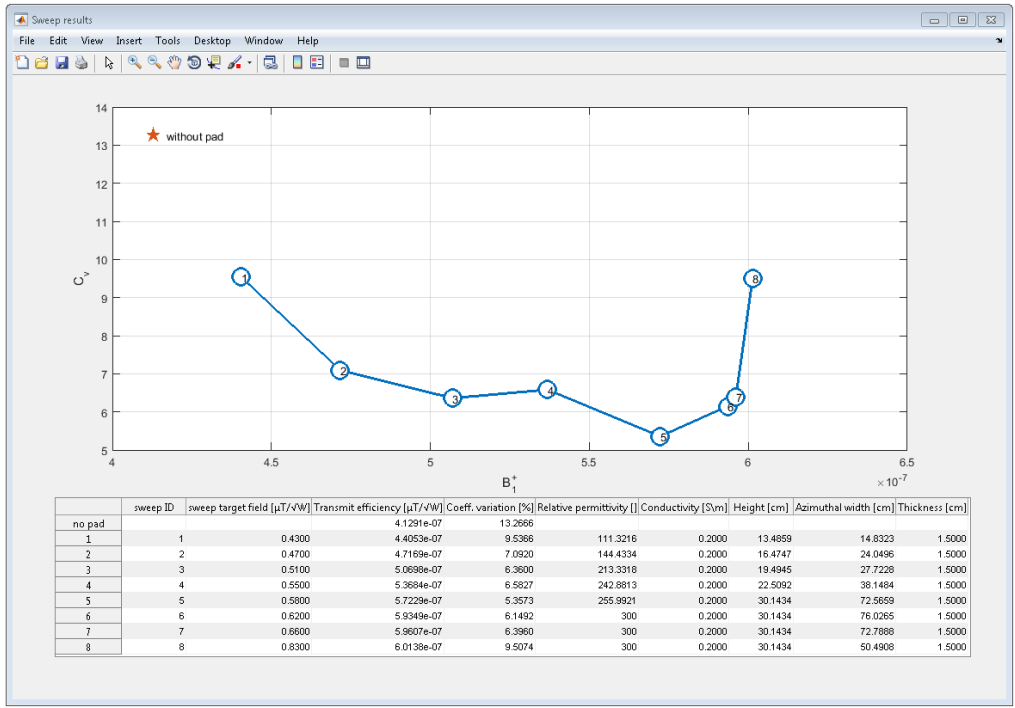

**(a)**

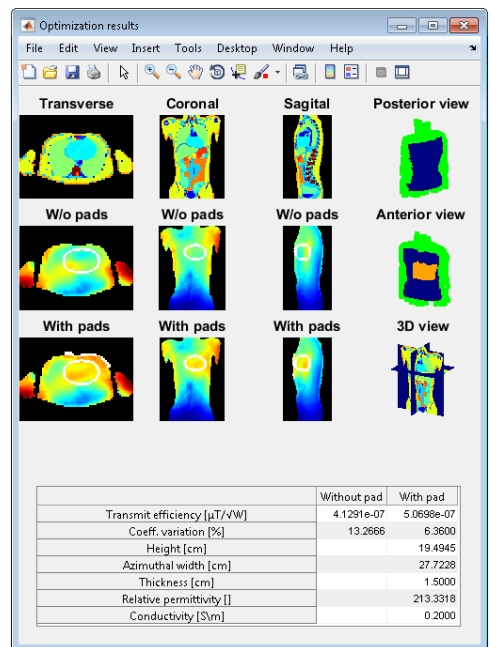

**(b)**

**Figure 4.** Design procedure for imaging the heart at 3T. In **(a)** the result is shown for a parameter sweep over a predefined set of target fields, after which a desired design can be selected based on the efficiency-homogeneity plot. For the selected design (here, number 3), a single optimization is performed to obtain the results as shown in **(b)** for later reference. Design number 5 is not chosen here, as the dimensions of the dielectric pad are not practical.

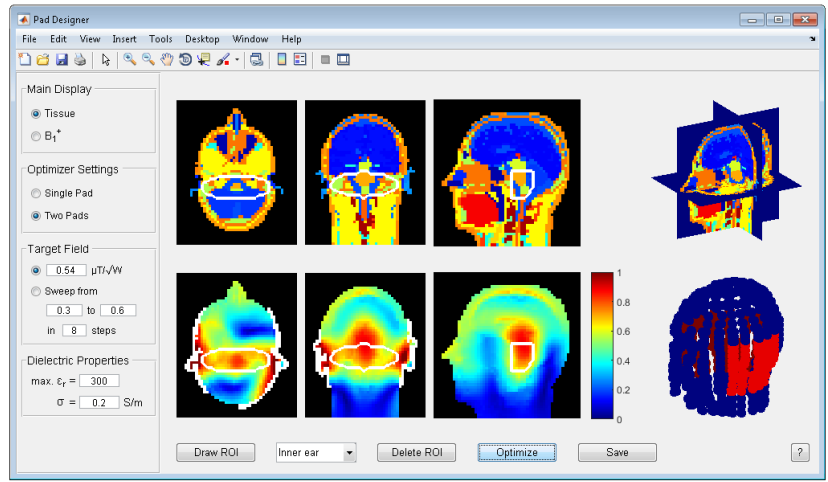

**(a)**

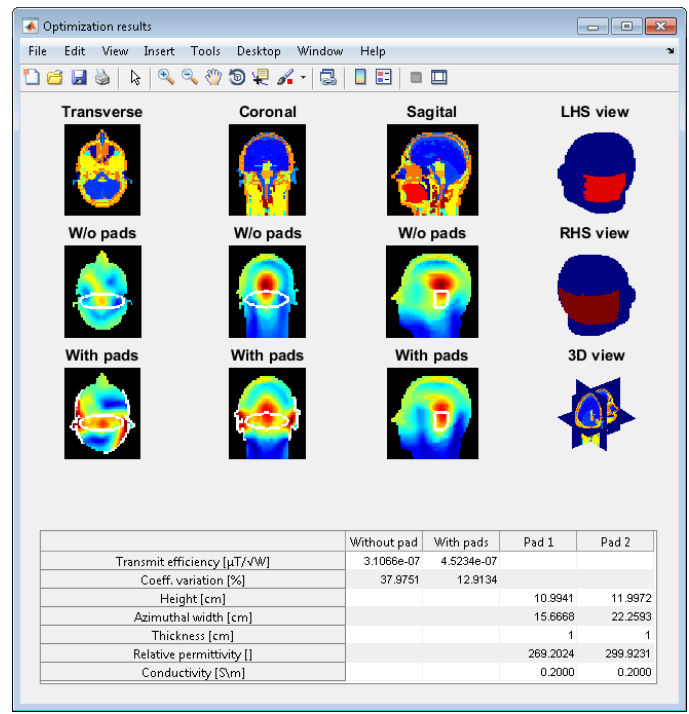

**(b)**

**Figure 5.** Pad design results for the inner ear using two dielectric pads **(a)**. A summary of the results is shown in **(b)**, where the improvement of the dielectric on the transmit efficiency and the coefficient of variation is shown. The dimensions and constitution of the two dielectrics are listed as well.